\documentclass[journal = nalefd]{achemso}
\setkeys{acs}{keywords = true}

\usepackage{graphicx}
\usepackage{amsmath}

\usepackage[pdftex,pdfauthor={Hiram Conley},pdftitle={Bandgap Engineering of Strained Monolayer and Bilayer MoS2},pdfsubject={Letter}, pdfkeywords={MoS2, strain engineering, band structure}, pdfproducer={Latex with hyperref}, pdfcreator={pdflatex}]{hyperref}

\author{Hiram J. Conley}
\author{Bin Wang}
\author{Jed I. Ziegler}
\author{Richard F. Haglund Jr.}
\affiliation{Department of Physics and Astronomy, Vanderbilt University}
\author{Sokrates T. Pantelides}
\affiliation{Department of Physics and Astronomy, Vanderbilt University}
\alsoaffiliation{Materials Science and Technology Division, Oak Ridge National Laboratory}
\author{Kirill I. Bolotin}
\affiliation{Department of Physics and Astronomy, Vanderbilt University}
\email{kirill.bolotin@vanderbilt.edu}

\title{Bandgap Engineering of Strained Monolayer and Bilayer MoS$_2$}

\begin{document}


\begin{abstract}

We report the influence of uniaxial tensile mechanical strain in the range 0--2.2\% on the phonon spectra and bandstructures of monolayer and bilayer molybdenum disulfide (MoS$_2$) two-dimensional crystals. First, we employ Raman spectroscopy to observe phonon softening with increased strain, breaking the degeneracy in the $E'$ Raman mode of MoS$_2$, and extract a Gr\"{u}neisen parameter of $\sim$1.06. Second, using photoluminescence spectroscopy we measure a decrease in the optical band gap of MoS$_2$ that is approximately linear with strain, $\sim$45 meV/\% strain for monolayer MoS$_2$  and $\sim$120 meV/\% strain for bilayer MoS$_2$. Third, we observe a pronounced strain-induced decrease in the photoluminescence intensity of monolayer MoS$_2$ that is indicative of the direct-to-indirect transition of the character of the optical band gap of this material at applied strain of $\sim$1\%. These observations constitute the first demonstration of strain engineering the band structure in the emergent class of two-dimensional crystals, transition-metal dichalcogenides.

\end{abstract}




Monolayer\footnote{Monolayer in this paper refers to one \textit{molecular} molybdenum disulfide (MoS$_2$) layer, or one layer of molybdenum atoms sandwiched between two layers of sulphur atoms. It is also sometimes refered to as trilayer MoS$_2$.} molybdenum disulfide (MoS$_2$),  along with other monolayer transition metal dichalcogenides (MoSe$_2$, WS$_2$, WSe$_2$) have recently been the focus of extensive research activity that follows the footsteps of graphene, a celebrated all-carbon cousin of MoS$_2$\cite{Wang2012}. Unlike semimetallic graphene, monolayer MoS$_2$ is a semiconductor with a large direct band gap\cite{Mak2010,Splendiani2010}. The presence of a band gap opens a realm of electronic and photonic possibilities that have not been previously exploited in two-dimensional crystals and allows fabrication of MoS$_2$ transistors with an on/off ratio exceeding $1\times10^8$ \cite{Radisavljevic2011,lin2012}, photodetectors with high responsivity\cite{Yin2012}, and even LEDs\cite{Sundaram2013}. Moreover, the direct nature of the band gap causes MoS$_2$ to exhibit photoluminescence at optical wavelengths\cite{Mak2010,Splendiani2010} with intensity that is tunable via electrical gating\cite{Newaz2013, Ross2013,Mak2013}. Finally, strong Coulomb interactions between electrons and holes excited across the band gap of MoS$_2$ lead to the formation of tightly bound excitons that strongly affect the optical properties of this material\cite{Ross2013,Mak2013}.  

It has been well established that straining a two dimensional material shifts its phonon modes, allowing a simple method to detect strain in these materials. These shifts, that are due to the anharmonicity of molecular potentials, can be probed with micro-Raman spectroscopy\cite{Huang2009,Mohiuddin2009}. Very recently, it has been proposed that mechanical strain can strongly perturb the band structure of MoS$_2$. It has been predicted that straining MoS$_2$ modifies the band gap energy and the carrier effective masses. Moreover, at strains larger than 1\% the lowest lying band gap changes from direct to indirect.\cite{Lu2012,Pan2012,Yue2012,Li2012a,Scalise2012,Shi2013}. It has been suggested that strain engineering of the band structure of MoS$_2$ could be used to increase carrier mobility of MoS$_2$, to create tunable photonic devices and solar cells\cite{Feng2012}, and even to control the magnetic properties of MoS$_2$\cite{Lu2012,Pan2012}. While strain perturbs the band structure of all materials, two-dimensional materials such as MoS$_2$ can sustain strains greater than 11\%\cite{Bertolazzi2011}, allowing exceptional control of material properties by strain engineering.

\begin{figure}[t]
\includegraphics[width=8.6 cm]{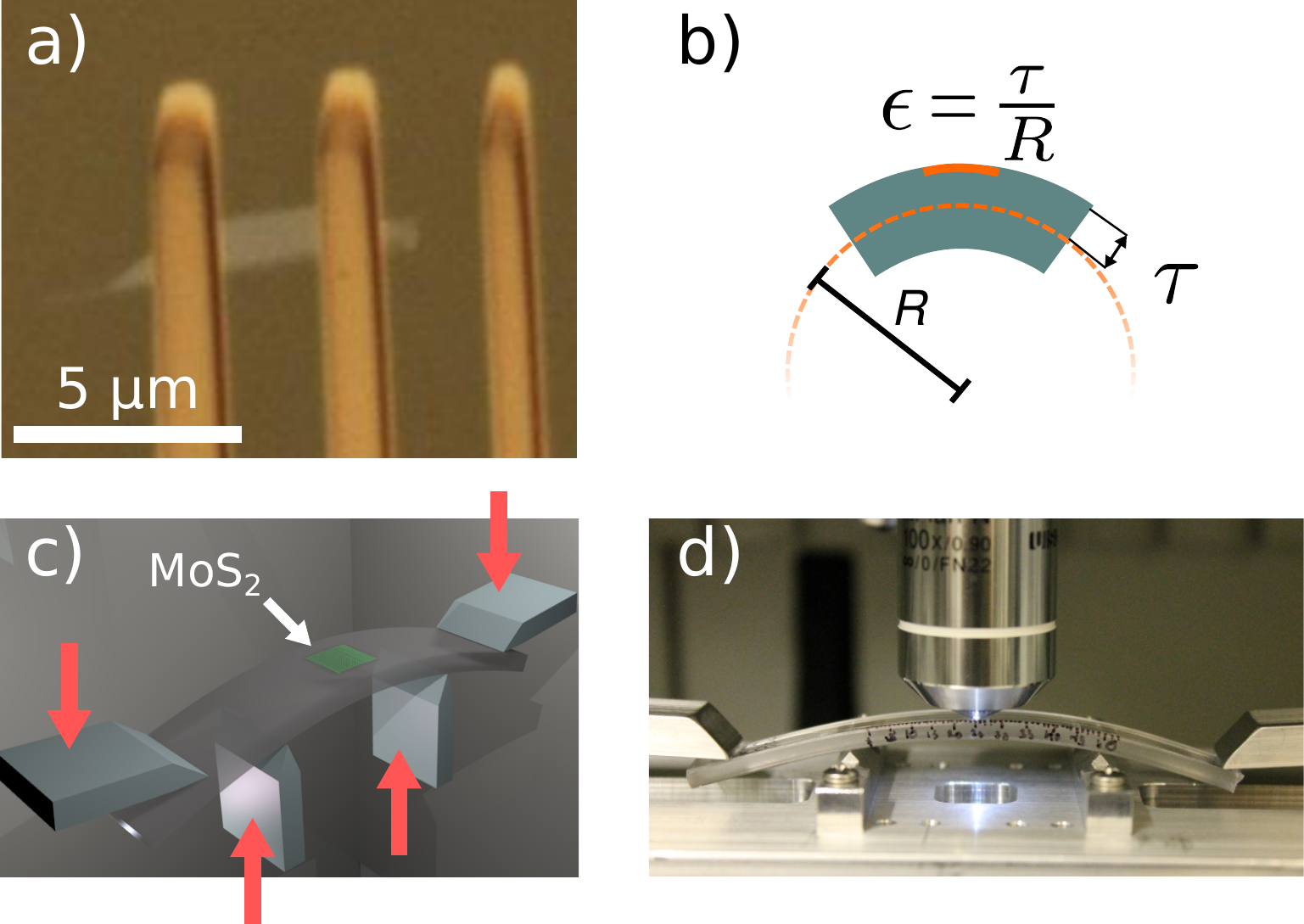}%
\caption{\textbf{Straining MoS$_2$ devices} (a) Optical image of a bilayer MoS$_2$ flake with titanium clamps attaching it to SU8/polycarbonate substrate. (b,c) Schematic of the beam bending apparatus used to strain MoS$_2$. (d) Photograph of bending apparatus with MoS$_2$ under strain.\label{fig_MoS2Image}}
\end{figure}

Here, we investigate the influence of uniaxial tensile strain from  0\% to 2.2\% on the phonon spectra and band gaps of both monolayer and bilayer MoS$_2$, by employing a four point bending apparatus (Fig. \ref{fig_MoS2Image}). First, with increasing strain, for both mono- and bilayer MoS$_2$ we observe splitting of the Raman peak due to the $E'$ phonon mode into two distinct peaks that shift by ~4.5 and ~1 cm$^{-1}$/\% strain. Second, a linear redshift of 45meV/\% strain of the position of the A peak in photoluminescence for monolayer MoS$_2$ (53meV/\% strain for bilayer MoS$_2$) indicates a corresponding reduction in band gap energy of these materials. Finally, we observe a pronounced strain-induced decrease in intensity of the photoluminescence of monolayer MoS$_2$. Our modelling and first-principles calculations indicate that this decrease is consistent with a transition of an optical band gap of MoS$_2$ from direct to indirect at $\sim$1\% strain, while the fundamental (or transport) band gap remains direct in the investigated regime of strain.

Fabrication of controllably strained devices starts by mechanically exfoliating\cite{Novoselov2005} MoS$_2$ onto a layer of cross-linked SU8 photoresist deposited onto a polycarbonate beam. The number of layers of MoS$_2$ is verified using Raman microscopy\cite{Li2012}. Titanium clamps are then evaporated through a shadow mask to prevent MoS$_2$ from slipping against the substrate (Fig. \ref{fig_MoS2Image}a). Uniaxial strain is applied to MoS$_2$ by controllably bending the polycarbonate beam in a four-point bending apparatus (Fig. \ref{fig_MoS2Image}c,d). Assuming that as-fabricated exfoliated devices before bending are virtually strain-free\cite{Chen2009}, we can calculate that upon bending the substrate with radius of curvature $R$, the induced strain in these devices is $\varepsilon=\tau/R$, where $2\tau=2$--3mm is the thickness of the substrate (Fig. \ref{fig_MoS2Image}b) \cite{Mohiuddin2009}. Overall, we fabricated four monolayer and three bilayer MoS$_2$ devices. The strained devices are probed with a confocal microscope (Thermo Scientific DXR) that is used to collect both Raman and photoluminescence spectra. We employ a 532nm laser excitation source with average power $\sim$100 $\mu$W, which does not damage our samples.

\begin{figure}[t]
\includegraphics[width=8.6 cm]{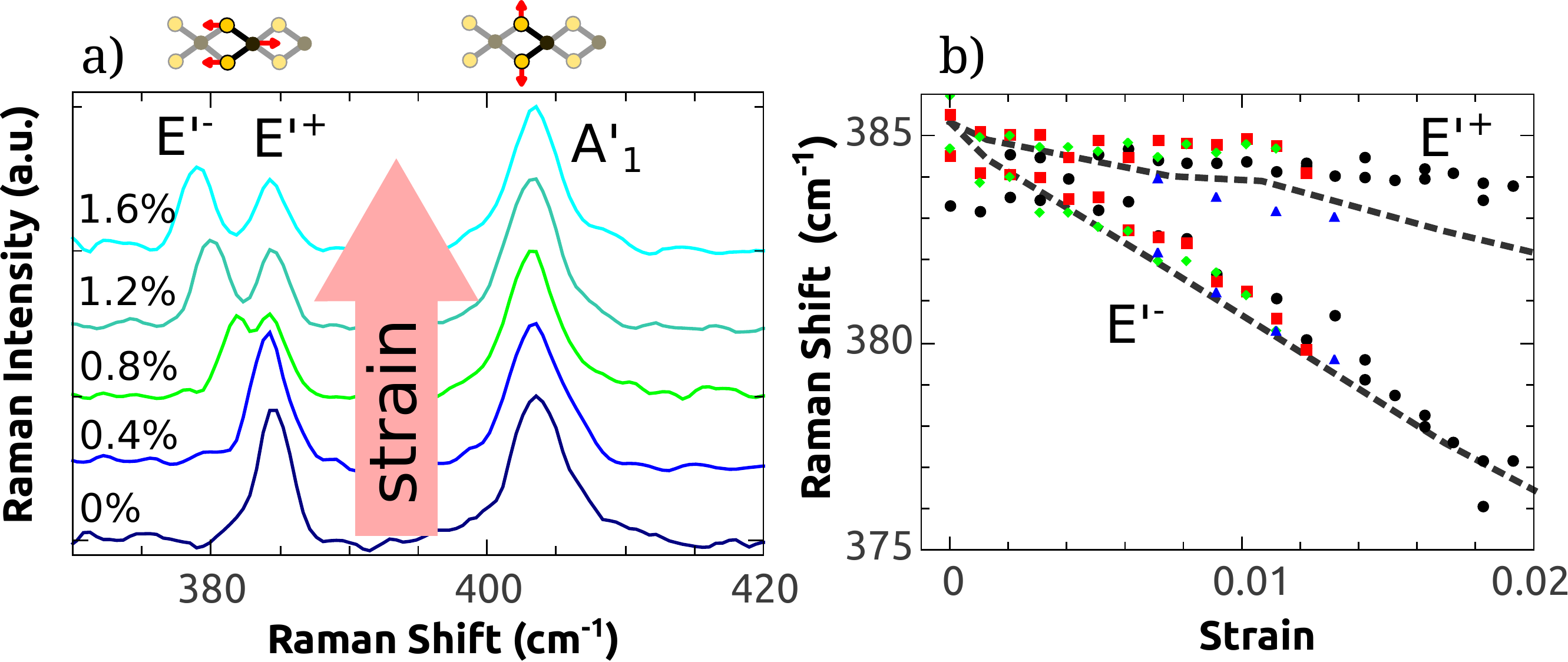}%
\caption{\textbf{Phonon softening of single layer MoS$_2$} (a) Evolution of the Raman spectrum as a device is strained from 0 to 1.6\%. (b) The peak location of the the $E'^+$ and $E'^-$ Raman modes, extracted by fitting the peaks to a Lorentzian, as their degeneracy is broken by straining MoS$_2$. Different colors represent individual devices. Dashed lines are the results of our first-principles calculations after subtraction of 9 cm$^{-1}$ to account for underestimating phonon energies. 
\label{fig_ramanshifts}}
\end{figure}

We first investigate the evolution of the Raman spectra of MoS$_2$ with strain (Fig. \ref{fig_ramanshifts}). In unstrained monolayer MoS$_2$  devices, consistent with previous reports\cite{Lee2010,Li2012}, we observe the $A'$ mode due to out-of-plane vibrations at 403 cm$^{-1}$ and the doubly degenerate $E'$ mode due to in-plane vibrations of the crystal at 384 cm$^{-1}$.

With increased strain, the $A'$ peak shows no measurable shift in position while the degenerate $E'$ peak splits into two subpeaks (in contrast to a previous report\cite{Rice2013}) that we label as $E'^{+}$ and $E'^{-}$ (Fig. \ref{fig_ramanshifts}), as strain breaks the symmetry of the crystal. The $A'$ mode maintains its intensity as strain increases, while the
total integrated intensity of the $E'$ peaks now splits between the
$E'^{+}$ and $E'^{-}$ peaks. The $E'^{-}$ peak shifts by 4.5 $\pm$ 0.3 cm$^{-1}/\%$ strain for monolayer devices and 4.6 $\pm$ 0.4 cm$^{-1}/\%$ strain for bilayer devices, while the $E'^{+}$ peak shifts by 1.0 $\pm$ 1 cm$^{-1}/\%$ strain for monolayer devices and 1.0 $\pm$ 0.9 cm$^{-1}/\%$ strain for bilayer devices, consistent with our first-principles calculations (dashed lines in Fig. \ref{fig_ramanshifts}; details of the calculations can be found in the supplementary materials). For applied strain in the range 0--2\%, the peak positions shift at nearly identical rates for all measured devices and do not exhibit hysteresis in multiple loading/unloading cycles, indicating that MoS$_2$ does not slip against the substrate and that the strain does not generate a significant number of defects. Bilayer devices behave in a similar manner but fail, either due to breaking or slipping of the MoS$_2$, at strains larger than 1\%.

The strain dependence of the Raman $E'$ mode enables us to calculate parameters characterizing anaharmonicity of molecular potentials, the Gr\"{u}neisen parameter, $\gamma$, and the shear deformation potential, $\beta$:
\begin{equation}
\gamma_{E'}=-\frac{\Delta\omega_{E'^+}+\Delta \omega_{E'^{-}}}{2 \omega_{E'} (1-\nu)\varepsilon}
\label{eq_GrunesianParameter}
\end{equation}
\begin{equation}
\beta_{E'}=\frac{\Delta\omega_{E'^{+}}-\Delta \omega_{E'^-}}{ \omega_{E'} (1+\nu)\varepsilon}
\label{eq_ShearDeformationPotential}
\end{equation} 
Here $\omega$ is the frequency of the Raman mode, $\Delta\omega$ is the change of frequency per unit strain, and $\nu$ is Poisson's ratio, which for a material adhering to a substrate is the Poisson ratio of the substrate, 0.33\cite{Mohiuddin2009}. This yields a Gr\"{u}neisen parameter of 1.1 $\pm$ 0.2,  half that of graphene (1.99)\cite{Mohiuddin2009} and comparable to that of hexagonal boron nitride (0.95--1.2)\cite{Kern1999,Sanjurjo1983}. The shear deformation potential is 0.78 $\pm$ 0.1 for both monolayer and bilayer MoS$_2$.

\begin{figure}[t]
\includegraphics[width=8.6 cm]{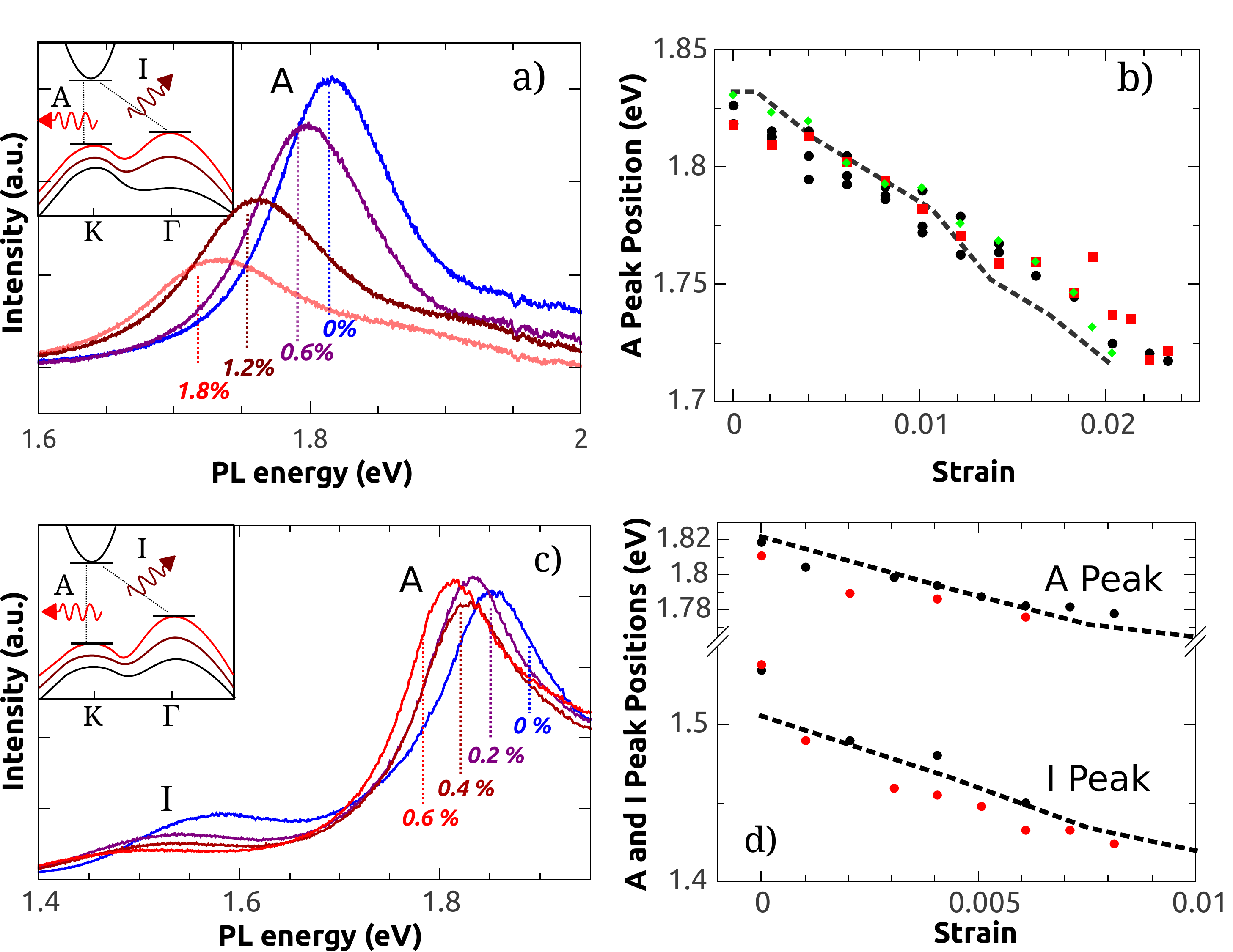}%
\caption{\textbf{Photoluminescence spectra of strained MoS$_2$} (a)  PL spectra of a representative monolayer device as it is strained from 0 to 1.8\%. Strain independent PL background was subtracted. (b) Evolution of the position of the A peak of the PL spectrum (Lorentzian fits) with strain for several monolayer devices (colors represent different devices) with GW$_0$-BSE calculations (dashed line) of expected peak position after 25 meV offset. (c) PL spectra of a representative bilayer device as strain is increased from 0 to 0.6\%. (d) PL peak position versus strain for the A and I peaks of bilayer devices (colors represent different devices) with good agreement to our GW$_0$-BSE calculations (dashed lines). Insets in (a) and (b) contains schematic representations of the band structure for both monolayer and bilayer MoS$_2$ devices that are progressively strained from 0\% (black) to $\sim$5\% (maroon) and $\sim$8\% (red). 
\label{fig_RedShift}}
\end{figure}

Next, we investigate the evolution of the band structure of MoS$_2$ with strain through photoluminescence (PL) spectroscopy. The principal PL peak (A peak) in unstrained direct-gap monolayer MoS$_2$ at 1.82$\pm$0.02 eV (Fig. \ref{fig_RedShift}a) is due to a direct transition at the $K$ point (Fig. \ref{fig_RedShift}a, inset)\cite{Splendiani2010,Mak2010}. The B peak, due to a direct transition between the conduction band and a lower lying valence band, is obscured in our devices by background PL of polycarbonate/SU8. The PL spectra of unstrained indirect-band gap bilayer MoS$_2$ devices are characterized by a similar A peak at 1.81$\pm0.02$ eV that originates from the same direct transition, but that is now less intense as it originates from hot luminescence.  In addition, we observe an I peak at 1.53$\pm0.03$eV (Fig. \ref{fig_RedShift}c), which originates from the transition across the indirect band gap of bilayer MoS$_2$ between the $\Gamma$ and $K$ points, (Fig. \ref{fig_RedShift}c, inset). 

\begin{figure}[t]
\includegraphics[width=8.6 cm]{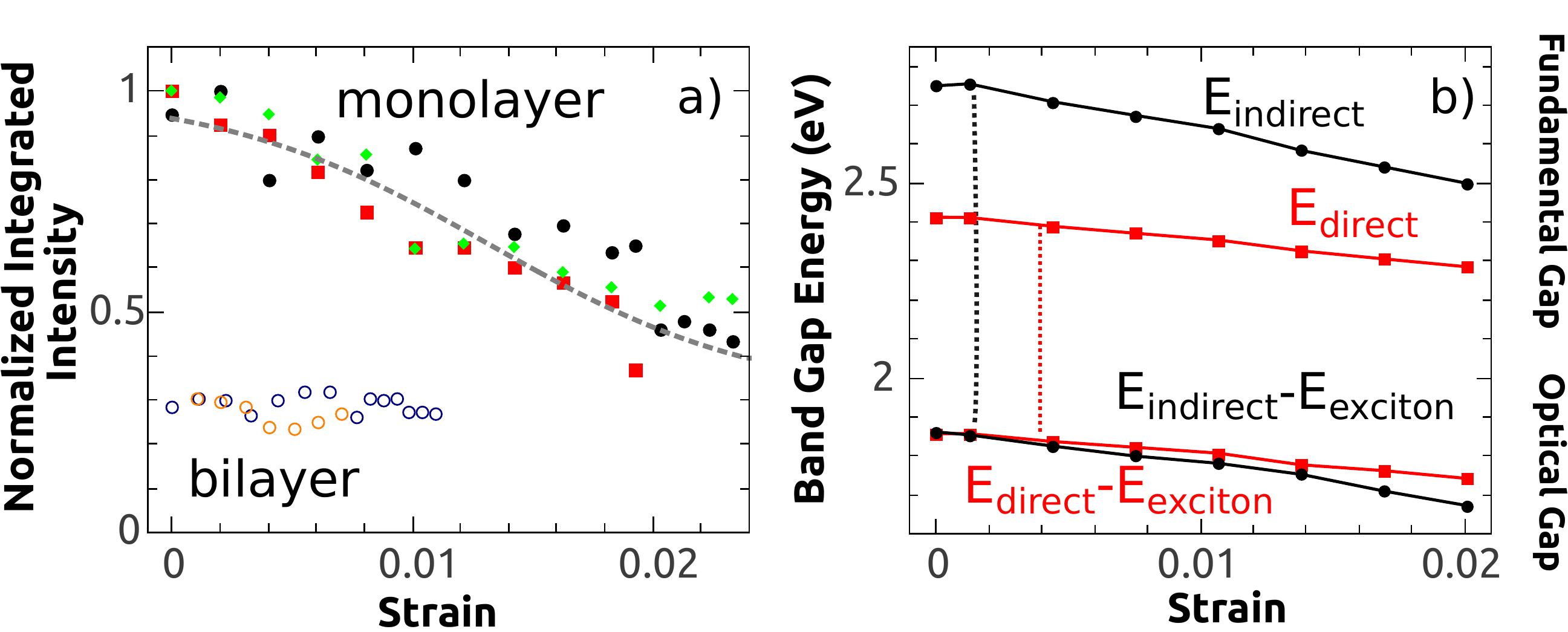}%
\caption{\textbf{Direct to indirect band gap transition in MoS$_2$} (a) Evolution of intensity of the A peak of strained monolayer MoS$_2$ (solid shapes) with a fit (dashed curve) to the rate equations consistent with a degenerate direct and indirect optical band gap at 1.3$\pm$0.6\% strain (supplementary material). PL intensity of bilayer A peak (unfilled circles) with no measurable change in intensity. Each color represents a distinct device. (b) GW$_0$ calculations of the fundamental band gaps of strained monolayer MoS$_2$, with an expected degeneracy at $\sim$5\% strain. Optical band gap calculated by including the exciton binding energy yields a degeneracy at $\sim$ 0.1\%.
\label{fig_BandGap}}
\end{figure}

Applied strain significantly changes the PL spectra (Fig. \ref{fig_RedShift}a,c). For all measured monolayer devices, the A peak redshifts approximately linearly with strain, at a rate of 45 $\pm$ 7 meV/\% strain (Fig. \ref{fig_RedShift}b). For bilayer devices, the A and I peaks are redshifted by 53 $\pm$ 10 and 129 $\pm$ 20 meV/\% strain respectively (Fig. \ref{fig_RedShift}d). While the intensity of the A peak in monolayer devices decreases to a third of its original size with an applied strain of 2\%, in bilayer devices the intensity of this peak is virtually strain-independent (Figs. \ref{fig_BandGap}a).

To understand our experimental results, we compare them to the results of GW$_0$-BSE calculations; details are given in supplementary materials. Crucially, these calculations capture the effect of strong electron-electron interactions in MoS$_2$ leading to the formation of excitons with binding energies significantly exceeding $k_BT$ at room temperature \cite{Cheiwchanchamnangij2012,Ramasubramaniam2012}. This is important because PL spectroscopy probes the optical band gap, the difference between the fundamental (or transport) band gap and the exciton binding energy.

The observed redshift of the PL peaks is indicative of strain-induced reduction of band gaps in both monolayer and bilayer MoS$_2$. Indeed, our GW$_0$-BSE calculations for a monolayer predict a reduction of the optical band gap at a rate of $\sim$59 meV/\% strain (dashed line in Fig. \ref{fig_RedShift}b), in close agreement with the measured PL peak shift. In bilayer devices, the calculated rates of reduction for the direct (67 meV/\% strain) and indirect (94 meV/\% strain) optical band gaps (dashed lines in Fig. \ref{fig_RedShift}d) are also in close agreement with measured redshift rates for A and I peaks, 53 $\pm$ 10 and 129 $\pm$ 20 meV/\% strain respectively.

We interpret the rapid decrease in PL intensity of monolayer MoS$_2$ with strain as a signature of the anticipated strain-induced transformation of the \textit{optical} band gap of this material from direct to indirect\cite{Lu2012,Pan2012,Shi2013,Feng2012}. Indeed, at zero strain the energy difference between the minimum of the conduction band at the $K$ point and the local maximum of the valence band at the $\Gamma$ point (the indirect gap) is higher in energy than the direct gap at the $K$ point (Fig. \ref{fig_RedShift}a, inset, black curve). However, we calculate that the indirect gap reduces with strain faster than the direct gap (59 vs. 94 meV/\% strain). As a result, if we ignore the effect of excitons, at $\varepsilon\sim$5\% the indirect gap overtakes the direct gap and monolayer MoS$_2$ becomes an indirect-gap material. With excitonic effects included, our calculations indicate that the direct and indirect \textit{optical} gaps (fundamental gaps minus binding energy of corresponding excitons) become degenerate at a much lower strain, 0.1\% (Fig. \ref{fig_BandGap}b). We however note that the accuracy of this value sensitively depends on the precise binding energy of the direct and indirect excitons that have not yet been measured experimentally.

As monolayer MoS$_2$ is strained and transitions from a direct to an indirect band gap material, we expect a marked decrease in the intensity of the A peak, as a majority of the excitons would not reside in this higher energy excitonic state, in good agreement the decrease in intensity in Fig. \ref{fig_BandGap}a. Quantitatively, a simple model describing direct and indirect excitons in monolayer MoS$_2$ as a two-level system yields an acceptable fit to our experimental data (dashed curve in Fig. \ref{fig_BandGap}a), with a direct-to-indirect band gap transition at $1.3\pm0.6$\% strain (details in supplementary information).

The observed PL spectra warrant two more comments. First, while for strains where the indirect band gap of monolayer MoS$_2$ is lower in energy than the direct band gap, we do not observe a peak corresponding to an indirect transition in its PL spectrum. This is likely due to the much smaller intensity of the indirect photoluminescence compared to the intensity of hot luminescence of the A peak. Second, in the range of strains from 1.3--5\%, monolayer MoS$_2$ enters a curious regime where its fundamental band gap is direct, while the optical band gap is indirect.

In conclusion, we have observed strain-induced phonon softening, band gap modulation and a transition from an optically direct to an optically indirect material in strained MoS$_2$ samples. These observations support a view of strain engineering as an enabling tool to both explore novel physics in MoS$_2$ (and other two-dimensional transition metal dichalcogenides such as MoSe$_2$, WS$_2$, WSe$_2$) and to tune its optical and electronic properties. An interesting avenue of research would be to explore the regime of degenerate direct and indirect bands -- that play key roles in a plethora of spin-related properties of the MoS$_2$\cite{Mak2012,Zeng2012}. Among the potential applications of the strain-dependent photoluminescence of MoS$_2$ and its cousins are nanoscale stress sensors and tunable photonic devices -- LEDs, photodetectors, and electro-optical modulators. 

Note: While the manuscript was under review, two studies reporting
modification of the band-gap with strain became available\cite{He2013,Wang2013}. The measured band gap shifts with strain are in general agreement with our results, however due to the much lower range of strains employed in the devices of Ref. [32,33] they do not measure the Gr\"{u}neisen parameter, probe the direct-to-indirect optical band gap transition, or observe contribution of excitonic effects.

\begin{acknowledgement}
This research was supported by NSF DMR-1056859, NSF EPS-1004083, and ONR N000141310299. BW and JZ were supported by DTRA HDTRA1-1-10-1-0047. We thank John Fellenstein for help in designing the four point bending apparatus, Branton Campbell for teaching us about phonon naming conventions, and Ashwin Ramasubramaniam for discussions about the first-principles calculations.
\end{acknowledgement}

Supporting Information Available: Supporting information, including bilayer Raman data, details of first principles calculations, and a simple two-level model to calculate the intensity of strained MoS$_2$, is available free of charge via the Internet at http://pubs.acs.org.

\bibliography{bibfile}

\end{document}